\documentclass[superscriptaddress,secnumarabic,twocolumn,
amssymb,amsmath,nobibnotes,aps,prd,showpacs,showkeys,nofootinbib]{revtex4}%
\usepackage{graphicx}
\usepackage{epsf}
\usepackage{bm}
\usepackage{amsmath}
\usepackage{amsfonts}
\usepackage{bigints}
\usepackage{amssymb}
\usepackage{epstopdf}
\usepackage{natbib}%
\providecommand{\U}[1]{\protect\rule{.1in}{.1in}}
\newcommand{\be}{\begin{equation}}
\newcommand{\ee}{\end{equation}}

\newcommand{\mincir}{\raise
-3.truept\hbox{\rlap{\hbox{$\sim$}}\raise4.truept\hbox{$<$}\ }}
\newcommand{\magcir}{\raise
-3.truept\hbox{\rlap{\hbox{$\sim$}}\raise4.truept\hbox{$>$}\ }}

\usepackage{color}

\begin{document}
\title{Exact solutions, finite time singularities and non-singular universe models from a variety of $\Lambda(t)$ cosmologies}

\author{Supriya Pan\footnote{Present Address: Department of Mathematics, Raiganj Surendranath Mahavidyalaya, Sudarshanpur, Raiganj, West Bengal 733134, India}}
\email{span@iiserkol.ac.in; span@research.jdvu.ac.in}
\affiliation{Department of Physical Sciences, Indian Institute of Science Education and Research Kolkata, Mohanpur 741246, West Bengal, India}

\keywords{Dynamical vacuum; Singularities; Nonsingular universe}
\pacs{98.80.-k, 95.36.+x}
\begin{abstract}
Cosmological models with time dependent $\Lambda$ (read as $\Lambda (t)$) have been investigated widely in the literature. Models that solve background dynamics analytically, are of special interest. Additionally, the allowance of past or future singularities at finite cosmic time in a specific model signals for a generic test on its viabilities  with the current observations. Following these,
in this work we consider a variety of $\Lambda (t)$ models focusing on their  evolutions and singular behaviour. We found that a series of models in this class can be exactly solved
when the background universe is described by a spatially flat Friedmann-Lema\^{\i}tre-Robertson-Walker (FLRW) line element. The solutions in terms of the scale factor of the FLRW universe offer different universe models, such as power law expansion, oscillating, and the singularity free universe as well. However, we also noticed  that a large number of models in this series permits past or future cosmological singularities at finite cosmic time. At last we close the work with a note that the avoidance of future singularities is possible for certain models under some specific restrictions. 

\end{abstract}

\maketitle
\section{Introduction}
\label{Intro}

Current observational data \cite{Ade:2015xua} provide a strong suggestion for an accelerating phase of the universe. This accelerating phase is generally believed to be driven by some hypothetical `dark energy' fluid. Now, in search of the dark energy fluid, the cosmological constant Lambda, $\Lambda$, is considered to be the simplest and elegenat dark energy candidate due to its extreme support with the available astronomical data. However, inspite of its great success, the cosmological constant problem has been seriously investigated in the last several years which thus instigated cosmologists to look for alternative ways in the names of dynamical dark energy models \cite{AT}, modified gravity theories \cite{Nojiri:2006ri, Nojiri:2010wj, DeFelice:2010aj,Sotiriou:2008rp,Capozziello:2011et, Cai:2015emx,Nojiri:2017ncd}. In a similar fashion, time dependent $\Lambda$ theory, where the
equation of state for Lambda is pegged at `$-1$', has also been attracted in the current cosmological research. In general such varying Lambda models can be considered to explore the expansion history, at least from the phenomenological ground. Interestingly, the class of cosmological models where $\Lambda \equiv \Lambda (t)$, can be viewed as the interacting dark energy models which are believed to be a potential class of cosmological models having an explanation to the cosmic coincidence problem\footnote{The cosmic coincidence is another generic problem in the dynamic dark energy models which states why the energy densities of dark energy and dark matter are of same order at present epoch?}. Thus, from this ground, cosmology with $\Lambda (t)$ might be considered to be an interesting theory to investigate the physics of dark energy as well as to focus on the dynamical history of the universe. However, the field equations for this theory do not follow from any covariant action. Nevertheless, in the last several years, a series of works towards this direction has been performed with several interesting results \cite{BC, CPBA, CDPA, PDCA, Basilakos:2009wi, Bauer:2009ea, Lima:2012mu, Perico:2013mna, Basilakos:2013xpa, Gomez-Valent:2014rxa, Gomez-Valent:2014fda, Gomez-Valent:2015pia,Oikonomou:2016pnq} and very recently another 
series of works \cite{rnv1,rnv2,rnv3,rnv4,rnv5, Kumar:2017dnp} showing that the current astronomical data from different independent sources favor the $\Lambda(t)$ class of models in compared to the concordance $\Lambda$CDM cosmology. In view of the observational data, $\Lambda(t)$ models could be an emerging class of cosmological models, for instance a suitable choice of some models belonging to this class can describe the entire cosmic histroy \cite{Lima:2012mu, Perico:2013mna,Basilakos:2013xpa}. To search for an appropriate time-dependent Lambda term, a promising development could be the use of renormalization group theoretical techniques in the quantum field theory (QFT) of curved space time which constrains the possible functional form for $\Lambda (t)$ \cite{Sola:2007sv,Shapiro:2009dh,Sola:2013gha,Sola:2015csa}, however, on the other hand, since QFT is not a completely developed subject yet, thus, in principle one can consider a wide ranges of models to probe the expansion history of the universe. 
 
In the current work, assuming
Friedmann-Lema\^{\i}tre-Robertson-Walker (FLRW) line element with zero spatial curvature as the best geometrical description for our universe, we consider a cosmlogical scenario with time dependent Lambda term. We show that such time varying Lambda cosmologies are equivalent to a cosmological scenario in terms of a generalized equation of state that can be derived from the generalized gravity theory \cite{Nojiri:2005sr, Nojiri:2006ri, Capozziello:2005pa,  Nojiri:2010wj} which includes the Einstein gravity as a special case, and hence it has a covariant formulation. Thus, following this, we consider a series of $\Lambda (t)$  models in such universe and found the exact evolution equations for the scale factor of the FLRW universe. 
Specifically, we find that the models are highly sensitive with their free parameters which often bring in singularities in the past or future evolution of the universe. For the singularities appearing in future, a classification depending on their nature can be found in \cite{Nojiri:2005sx}. However,
to the best of our knowledge the appearance of singularities in $\Lambda (t)$ cosmologies has not been focused much in the literature. Thus, in the present work we have tried to highlight the appearance of the cosmological singularities that are found to exist in a class of  $\Lambda (t)$ cosmologies.

The paper has been organized as follows: In section \ref{section2}, we shortly describe the background equations in a FLRW universe. Section \ref{sec3} describes the models considered in this work and their analytic solutions. Finally, in section \ref{discu} we draw our findings in short. 

We note that any subindex ``\textit{zero}'' attached to any quantity describes its value at present epoch.

\section{Field equations}
\label{section2}

As usual let us consider that our universe is well described by the spatially flat FLRW line element $ds^2= -dt^2+ a^2 (t)\left[dr^2 + r^2 \left( d\theta^2+ \sin^2 \theta \, d\phi^2 \right) \right]$, where $a(t)$ is the expansion scale factor of the universe. Further, we consider a mixture of perfect fluids with total energy-momentum tensor  $T^{\mu \nu}_{tot}= \sum _{i} T^{\mu\nu}_{i}$. Thus, 
from the conservation equation 
$\nabla _{\nu} T^{\mu \nu}_{tot}= 0$, 
one arrives at 

\begin{eqnarray}
\sum_{i} \dot{\rho}_i + 3 H \sum_{i} (p_i + \rho_i) = 0
\end{eqnarray}
where an overhead dot represents the cosmic time 
differentiation;
$p_i$, $\rho_i$, are respectively the pressure and energy density of the $i$-th
fluid; and  $H \equiv \dot{a}/a$, is the Hubble rate which for spatially flat FLRW 
universe becomes, $3 H^2 = \sum_{i} \rho_i$, in the units $8 \pi G = 1$. We consider that  the mixture of fluids comprises of a matter sector and the vacuum energy which is dynamical. For convenience, we denote $p$, $\rho$ as the pressure and the energy density of the matter while the dynamical vacuum is represented by $\Lambda (t)$. Following the conservation equation of the mixture of fluids described above, we find that  

\begin{eqnarray}\label{continuity}
\dot{\rho}+ 3\, H\, (\rho +p) = -\, \dot{\rho}_{\Lambda} = - \dot{\Lambda}(t)~.
\end{eqnarray}
Now one can write down the Friedmann equation explicitly as 
\begin{align}\label{Friedmann}
3H^2= \rho + \Lambda(t)~,
\end{align}
and using (\ref{continuity}), (\ref{Friedmann}) with the equation of state $p= (\gamma- 1)\, \rho$ (where $1 \leq \gamma \leq 2$), we get the following first order
differential equation

\begin{align}\label{main-eqn}
2 \dot{H} + \gamma \left[3 H^2 - \Lambda (t) \right] & = 0~,
\end{align}
which is the main differential equation to understand the dynamics of the universe when $\Lambda (t)$ is specified. The field equations (\ref{Friedmann}), (\ref{main-eqn}) as one may note, do not follow from any covariant action. This is a generic problem in the time varying Lambda cosmologies while so many interesting results follow from this theory. However, interestingly enough, one can show that an equivalent description of the above field equations can be reproduced from a covariant action that leads to an inhomogeneous equation of state. Such inhomogeneous equation of state can be derived from the $f(R)$-gavity \cite{Nojiri:2005sr, Nojiri:2006ri, Capozziello:2005pa,  Nojiri:2010wj} or scalar-tensor theory \cite{Capozziello:2005pa} or even from the braneworld scenario \cite{Nojiri:2005sr}. Here we recall the $f(R)$-gravity models with the action

\begin{eqnarray}
S = \int \sqrt{-g}\, d^4 x \left[ \frac{R}{2} + f(R) + \mathcal{L}_m \right] 
\end{eqnarray} 
where $f(R)$ is any arbitrary function of the scalar curvature $R$ and $\mathcal{L}_m$ is the largangian for the matter sector. Now, for the FLRW universe, the field equations can be written as  \cite{Nojiri:2005sr, Nojiri:2006ri, Capozziello:2005pa, Nojiri:2010wj} 

\begin{eqnarray}
3 H^2 = \rho + \rho_{\texttt{grav}},\\
2 \dot{H} + 3 H ^2 = - p - p_{\texttt{grav}}
\end{eqnarray}
where $\rho$, $p$ are respectively the energy density and the pressure of the matter field while $\rho_{\texttt{grav}}$, $p_{\texttt{grav}}$ are respectively the geometrical energy density and the pressure appearing due to the modifications in the gravity sector with the forms given by \cite{Nojiri:2005sr, Nojiri:2006ri, Capozziello:2005pa, Nojiri:2010wj}

\begin{eqnarray}
\rho_{\texttt{grav}} = 6 \left( \dot{H} + H^2- H \frac{d}{dt} \right) f'(R),\\
p_{\texttt{grav}} = f(R) - 2 \left[ \dot{H} + 3 H^2 + \frac{d^2}{dt^2} + 2 H \frac{d}{dt}\right] f' (R),
\end{eqnarray}
where the prime stands for the differentiation with respect to the scalar curvature $R$. Now, introducing, $\rho_t = \rho+ \rho_{\texttt{grav}}$, as the total energy density or effective energy density and similarly $p_t = p+ p_{\texttt{grav}}$ as the total or effective pressure of the system, and then following \cite{Nojiri:2005sr, Nojiri:2006ri, Capozziello:2005pa, Nojiri:2010wj}  one can easily come up with an inhomogeneous equation of state taking the form

\begin{eqnarray}\label{eqn-new}
p_{t} = (\gamma- 1) \rho_t + \gamma f(R)- \Bigg[ (3\gamma- 2) \dot{H} + 3 \gamma H^2 \nonumber\\- \frac{d^2}{dt^2} - (3\gamma -1) H \frac{d}{dt} \Bigg] f' (R)
\end{eqnarray}
where recall that the equation of state for the matter sector is assumed to be $p = (\gamma -1) \rho$ and here $R = 6 \dot{H} + 12 H^2$. It is evident that the last two terms of the equation (\ref{eqn-new}) induce a function of the Hubble rate and its higher derivatives, that means, 
the equation (\ref{eqn-new}) looks like 
\begin{eqnarray}\label{eqn-new1}
p_t = (\gamma -1)\rho_t + F (H, \dot{H}, \ddot{H},...). 
\end{eqnarray}
The dynamics and 
evolution of the cosmological models with certain choices for $F (H, \dot{H}, \ddot{H},...)$ have been discussed in \cite{Nojiri:2005sr, Nojiri:2006ri, Capozziello:2005pa, Nojiri:2010wj}. 
One can see that the equation (\ref{eqn-new}) for 
$f(R) = c R$ ($c$ being a constant), can be recast as \cite{Nojiri:2006ri, Capozziello:2005pa}

\begin{eqnarray}\label{eqn-new2}
p_{t} = (\gamma -1)\rho_{t} + 6 c \dot{H} + 6 \gamma c H^2~.
\end{eqnarray}

Now, in presence of the time varying Lambda models, one can see that, $\rho_{t} = \rho+\rho_{\Lambda}$, as the total energy density of the universe and $p_{t} = p+ p_{\Lambda}$, as the total pressure. Using these two one arrives at,
\begin{eqnarray}\label{eqn-new3}
p_t = (\gamma -1 )\rho_t - \gamma \Lambda (t),
\end{eqnarray}
which resembles with the equation (\ref{eqn-new1}). And in particular, comparison of (\ref{eqn-new2}) and (\ref{eqn-new3}) gives $\Lambda (t) = -6 c \gamma^{-1} (\dot{H} + \gamma H^2)$. Thus, we find that the time varying Lambda cosmologies can be looked as a cosmological scenario with inhomogeneous equation of state where the gravity could be described by either general relativity, modified $f(R)$ gravity, or even the braneworld gravity. Now, one can pick up different functional forms for $\Lambda (t)$ either in terms of the cosmic time or in terms of the Hubble rate as well as its time-derivatives, and solve the  master equation (\ref{main-eqn}).

Now, there is another interesting observation that we would like to remark here. Actually, such observation is very helpful to solve the master equation (\ref{main-eqn}) in a different way. In fact, it is an indirect method to solve the master equation (\ref{main-eqn}) without using $\Lambda (t)$. To do so let us recall some previous equations, $\rho_{t} = \rho+\rho_{\Lambda}$, $p_{t} = p+ p_{\Lambda}$. With these, one can recast the Friedmann equation as 
$3 H^2 = \rho_{t}$,
and the conservation equation similarly 
becomes $\dot{\rho}_{t} + 3 H (\rho_{t}+p_{t}) = 0$. Taking a time derivative of the Friedmann equation and then using the conservation equation, one arrives at 
\begin{eqnarray}\label{sp-new}
2 \dot{H} = - 3H^2 -  p_{t},
\end{eqnarray}
which is like equation (\ref{main-eqn}) but without going via $\Lambda (t)$ approach.  Now the equation (\ref{sp-new}) under the transformation $y = 1/H$,  turns out to be 

\begin{eqnarray}
2 \frac{dy}{dt} = 3 + p_{t}\, y^2,
\end{eqnarray} 
which can be solved for $y\, (\,= 1/H)$, for various choices for $p_t$, and hence for the scale factor of the FLRW universe. However, the choices can be made for the quantity $p_t \, y^2$ too instead of $p_t$. One may realize that the problem reduces to the classfication of solutions to the integral 
\begin{eqnarray}
\int \frac{dy}{3+p_t\, y^2} = \frac{1}{2} \left(t + t_c \right)~,
\end{eqnarray}
where $t_c$ is the constant of integration. We note that here we do not specify any equation of state parameter unlike in previous approach, see eqn. (\ref{continuity}).

In this work we shall consider the explicit forms for $\Lambda (H)$ and discuss the cosmological solutions. 
In what follows, the next section is devoted to investigate such cosmological models \footnote{In this discussions, perhaps it will be interesting to note that if the expansion scale factor is gievn, then the evolution of $\Lambda (t)$ is straightforward. For instance, if one considers the power law cosmology: $a= a_0 \left( \frac{t}{t_0}\right)^n$, then $\Lambda (t) \propto H^2$, or alternatively, $\Lambda (t) \propto \dot{H}$.}. 

\section{$\Lambda(t)$ cosmologies}
\label{sec3}

In general, there is no bound to choose some particular choices
for $\Lambda (t)$ or $\Lambda (H)$ and solve the differential equation (\ref{main-eqn})
to obtain the cosmological solutions. Finally, one can use the observational data to test the viability of such models.  Thus, following this approach, we first recall a very natural choice for $\Lambda (H)$  as $\Lambda (H)= \sigma\, H$, where $\sigma >0$ is any constant with dimension equal to the dimension of the Hubble rate. This choice has been  studied earlier \cite{BC, CPBA, CDPA, PDCA} in order to account for the current expanding accelerating universe.
However, a series of analysis \cite{Gomez-Valent:2014rxa, Gomez-Valent:2014fda, Gomez-Valent:2015pia} reports that this linear model is strongly disfavored by the observational data, hence, the linear model is not of much interest for current study. The second possibility, an extension of the linear model, is the quadradic function in the Hubble parameter,  $\Lambda(H)= \bar{\sigma} H^2$ ($\bar{\sigma} > 0$). However, the authors in Refs. \cite{Basilakos:2009wi} (also see \cite{Gomez-Valent:2014rxa}) have established that this model suffers from several theoretical probelms as well as with the observational data. Thus, we see that the models $\Lambda (H) \propto H$ and $\Lambda (H) \propto H^2$, are not compatible with the current cosmology.  Therefore, a natural extension is to consider the linear combinations of $H$, $H^2$,... and the derivatives of of the Hubble parameter.  
In the following we introduce some general models of such kind and discuss the cosmological solutions and the dynamics of the universe.

\subsection{Model 1: $\Lambda (H)= \lambda+ \alpha\, H+ 3 \, \beta\, H^2$}
\label{new-section-number}

We recall a generalized model $\Lambda (H)= \lambda+ \alpha\, H+ 3 \, \beta\, H^2$, where $\lambda$, $\alpha$, $\beta$ are real constants. This model was first introduced in Ref. \cite{Basilakos:2009wi} and later on it was investigated in detail in Refs. \cite{Gomez-Valent:2014rxa, Lima:2012mu, Gomez-Valent:2014fda} in presence of the observational data where the authors found that the models with $\lambda =0$ have serious problems. They do not fit with the observational data and also with the linear growth rate of clustering. On the other hand, the models with $\lambda \neq 0$ could survive with the observational data and at late time they behave like $\Lambda$CDM cosmology. We noticed that for $\lambda \neq 0$, three different  cosmological solutions can be obatained in which only one was was discussed in \cite{Basilakos:2009wi}. Keeping aside the observational viabilities of the other two solutons, here we are interested with the theoretical implications.  For completeness, we shall discuss all the three possible solutions emerging from the above $\Lambda (H)$ model with $\lambda \neq 0$. 
Thus,  using the master differential equation (\ref{main-eqn}) for  $\Lambda (H)= \lambda+ \alpha\, H+ 3 \, \beta\, H^2$, we see

\begin{align}\label{main-eqn1}
2\, \dot{H} + \gamma\, \left( 3H^2-\lambda-\alpha\, H- 3\beta H^2 \right) & = 0.
\end{align}
Now, the differential equation (\ref{main-eqn1}) opens several possibilities depending
on its free parameters $\lambda$, $\alpha$, $\beta$. We shall be interested only the cosmological solutions for  $\beta \neq 1$, since it has been observed that the astronomical data do not favour the model with $\beta = 1$, see \cite{Lima:2012mu, Gomez-Valent:2014fda}. In the following we discuss three separate possibilities and hence three different cosmological solutions for the model $\Lambda (H)= \lambda+ \alpha\, H+ 3 \, \beta\, H^2$. Thus, we wish to solve the following integral 

\begin{align}
\int \frac{dH}{\left( H- \frac{\alpha}{6 (1-\beta)}\right)^2 + \frac{12 \lambda (\beta -1) -\alpha^2}{36 (1-\beta)^2}} = -\frac{3\gamma}{2} (1-\beta) + c~,
\end{align}
where $c$ is the constant of integration. In the following we shall investigate the cosmological solutions depending on the nature of the factor $\Delta = 12 \lambda (\beta -1) -\alpha^2$. \\

1. We consider the first possibility when $\Delta= 12 \lambda (\beta -1)- \alpha^2 < 0$. This has been discussed in \cite{Basilakos:2009wi} with the solutions for the Hubble factor and the scale factor as 

\begin{eqnarray}
H = \frac{1}{{6(1-\beta)}}\Bigg[ \alpha + B \tanh\Bigg(\frac{\gamma B}{4}(t-t_0)\nonumber\\ + \tanh^{-1} \left( \frac{6H_0(1-\beta)- \alpha}{B}\right)\Bigg) \Bigg],\label{sol2a}
\end{eqnarray}
where $B = \sqrt{\alpha^2- 12\lambda (\beta-1)}$. Now, the scale factor can be solved as 

\begin{align}\label{sol2b}
\frac{a}{a_0}= \exp\left(\frac{\alpha}{6(1-\beta)}(t-t_0)\right)\left[\frac{\cosh\left(\frac{\gamma}{4B} (t-t_0)+ A\right)}{\cosh A}\right]^F
\end{align}
where $F= \frac{2}{3\gamma (1-\beta)}$,  $A= \tanh^{-1} \left(\frac{6H_0 (1-\beta)-\alpha}{B}  \right)$.
The above solution of the scale factor represents a singularity free solution. \textit{Now, there are two more possibilities on $\Delta$, that we going to present below.} \newline

2. When $\Delta= 12 \lambda (\beta -1)- \alpha^2 = 0$. In that case, the sign of $\alpha$ does not matter, but we have two constraints 
over the other two parameters $-$ either $\lambda> 0,~\beta> 1$, or $\lambda< 0,~\beta< 1$. The solution for the Hubble rate and the scale factor becomes

\begin{eqnarray}\label{sol3a}
H = \frac{1}{6(1-\beta)} \left[\alpha + \frac{1}{\frac{\gamma}{4}(t-t_0)+ X}\right],
\end{eqnarray}
where $X= \frac{1}{6H_0 (1-\beta)-\alpha}$. The scale factor in this case becomes

\begin{align}\label{sol3b}
\frac{a}{a_0}=\exp\left(\frac{\alpha}{6(1-\beta)}(t-t_0)\right)\, \Bigg| \frac{\frac{\gamma}{4}(t-t_0)+ X}{X} \Bigg|.
\end{align}
Here, from eqn. (\ref{sol3b}), it is readily seen that the scale factor can be zero at some finite  cosmic time. It happens for $\frac{\gamma}{4} (t-t_0)+ X = 0$, that means we have

\begin{align}
t & = t_0 - \frac{4X}{\gamma} = t_0 -\frac{4}{\gamma}\, \frac{1}{6H_0 (1-\beta)- \alpha},
\end{align}
which further means we have a past singularity at finite time if $6 H_0 (1-\beta)-\alpha > 0$,
and on the other hand, we get a future singularity for $6 H_0 (1-\beta)-\alpha < 0$. Now, since the parameters $\alpha$, $\beta$ are at hand, thus, one can see that under certain restrictions over them, the future singularity can be avoided, however, the restriction on the parameters from observational grounds is a different issue which is beyond the scope of the current work. \newline

3. Finally, when $\Delta= 12 \lambda (\beta -1)- \alpha^2 > 0$, Hubble parameter is found to be

\begin{eqnarray}\label{sol4a}
H  = \frac{1}{1-\beta}\Bigg[\alpha + Y \tan\Bigg(-\frac{Y\gamma}{4}(t-t_0)\nonumber\\+ \tan^{-1}\left(\frac{6 H_0 (1-\beta)-\alpha}{Y}\right)\Bigg)\Bigg]
\end{eqnarray}
where $Y= \sqrt{12\lambda (\beta-1)- \alpha^2}$. Now, the scale factor is solved as

\begin{align}\label{sol4b}
\frac{a}{a_0} = \exp\left(\frac{\alpha}{6(1-\beta)} (t-t_0)\right) \Bigg| \frac{\cos \left( -\frac{Y\gamma}{4} (t-t_0) +\theta\right)}{\cos \theta} \Bigg|^G
\end{align}
where $G= \frac{2}{3\gamma(1-\beta)}$, $\theta= \tan^{-1} \left(\frac{6H_0 (1-\beta)-\alpha}{Y} \right)$. One can see that $\cos \theta \neq 0$. Thus, for $G >0$ (equivalently, $\beta <1$) the scale factor exhibits an oscillating nature. On the other hand, for $G <0$ (i.e. $\beta >1$), we find that the scale factor could diverge at finite time given by 

\begin{align}
t & = t_0 + \frac{4}{\gamma Y} \left[ \tan^{-1} \left(\frac{6H_0 (1-\beta)-\alpha}{Y} \right) - (2n+1)\, \frac{\pi}{2} \right],
\end{align}
where $n \in \mathbb{Z}$. The divergence in the scale factor may appear in the finite past or future 
depending on the sign of $\left[ \tan^{-1} \left(\frac{6H_0 (1-\beta)-\alpha}{Y} \right) - (2n+1)\, \frac{\pi}{2} \right]$. It is
clear that if $\left[ \tan^{-1} \left(\frac{6H_0 (1-\beta)-\alpha}{Y} \right) - (2n+1)\, \frac{\pi}{2} \right] < 0$, then the scale factor may diverge at finite time in the past while on the other hand, if $\left[ \tan^{-1} \left(\frac{6H_0 (1-\beta)-\alpha}{Y} \right) - (2n+1)\, \frac{\pi}{2} \right]> 0$, the scale factor could diverge at finite time in future.

Thus, we find that for both $\Delta > 0$ and $\Delta = 0$, the model allows finite time singularities. Now in the following we consider a very general model and discuss its cosmological solutions.

\subsubsection{The Model $\Lambda (H) = \beta H + 3 H^2 + \delta H^n$, $n \in \mathbb{R}- \{0, 1\}$}

We consider another model where $\Lambda (H) = \beta H + 3 H^2 + \delta H^n$, where $\beta, \delta \in \mathbb{R}$ and $n \in \mathbb{R}- \{0, 1\}$. The restriction on $n$ has been taken for a great purpose. For $n = 0$ or $1$, one can see that the model returns to two subcases of the model discussed in section \ref{new-section-number}.  Thus, for $n \in \mathbb{R} -\{0, 1\}$, one can include various powers for $H$ and hence a class of distinct cosmological scenarios. Since we already established an equivalence between the $\Lambda (t)$ cosmologies and the cosmology with a generalized (inhomogeneous) equation of state in section \ref{section2}, thus, one may connect with a similar model \footnote{One can see the model in equation (176)  or equation (3.137) of \cite{Bamba:2012cp}. } studied in \cite{Bamba:2012cp}. We also refer to an earlier work \cite{Barrow:1990vx} where this type of models produce inflationary scenario leading to power law or exponential universes along with intermediate inflationary solutions. The differential equation (\ref{main-eqn}) for this model becomes

\begin{eqnarray}\label{bernoulli}
\dot{H} - \frac{\gamma \beta}{2} H = \frac{\gamma \delta}{2} H^n
\end{eqnarray}
which is nothing but the Bernoulli's equation. Now one can solve the equation (\ref{bernoulli}) leading to the exact solution of the Hubble rate as

\begin{eqnarray}\label{bernoulli1}
H = \Bigg[ \left(H_0^{1-n} + \frac{\delta}{\beta} \right) \exp\left( \frac{\gamma \beta}{2} (1-n)  (t-t_0) \right) -\frac{\delta}{\beta} \Bigg]^{\frac{1}{1-n}}
\end{eqnarray}
where $\beta \neq 0$. Later we shall discuss the solutions for $\beta =0$. One may notice that for $\delta/\beta \lll 1$, the equation (\ref{bernoulli1}) becomes,
$H \simeq H_0 \exp \left(\frac{\gamma \beta}{2} (t-t_0) \right)$, and the scale factor becomes double exponential type.  However, without imposing any restriction on $\beta$ and $\delta$, the direct integration of (\ref{bernoulli1}) gives the scale factor as

\begin{align}\label{bernoulli1.1}
a = a_0 e^{\bigintss_{t_0}^{t} \Bigg[ \left(H_0^{1-n} + \frac{\delta}{\beta} \right) \exp\left( \frac{\gamma \beta}{2} (1-n)  (t-t_0) \right) -\frac{\delta}{\beta} \Bigg]^{\frac{1}{1-n}} dt}.
\end{align}
Now, if $n<1$ then the scale factor does not show any finite time singularities, but for $n>1$, the scale factor may encounter with finite time singularities. That is because of  the factor $1/(1-n)$. Certainly, for $n> 1$, solving the equation 
$$ \left(H_0^{1-n} + \frac{\delta}{\beta} \right) \exp\left( \frac{\gamma \beta}{2} (1-n)  (t-t_0) \right) -\frac{\delta}{\beta} = 0,$$
one finds that the divergence in the scale factor may appear for

\begin{eqnarray}\label{bernoulli1.2}
t = t_0 + \frac{2}{(n-1)\gamma \beta } \ln \Bigg| \frac{H_0^{1-n}+ \delta/\beta}{\delta/\beta}\Bigg|,\nonumber\\
= t_0 + \frac{2}{(n-1)\gamma \beta } \ln \Bigg|1+ \frac{\beta}{\delta}H_0^{1-n} \Bigg| 
\end{eqnarray}   
where the quantity $(H_0^{1-n}+ \delta/\beta )$ is non-zero \footnote{If we assume that $H_0^{1-n}+ \delta/\beta = 0$, that means one of $\delta$ or $\beta$ is negative, and in that case, from equation (\ref{bernoulli}), one finds an exponential expansion of the universe. }.  Now let us make a crucial investigation from (\ref{bernoulli1.2}) as follows. 
\begin{itemize}

\item We assume that $\delta, \beta >0$. In that case,  equation (\ref{bernoulli1.2}) directly shows that, $t> t_0$, which means that the scale factor diverges at finite future time. 

\item Now, if $\delta, \beta  < 0$. In that case, from equation (\ref{bernoulli1.2}) one finds that, $t < t_0$, which means that the scale factor diverges at past at finite time which is not physical.

\item For $\beta >0$, $\delta <0$, the scale factor diverges in past at finite time which is unphysical. 

\item For $\beta <0$, $\delta >0$, the scale factor diverges at finite future time. 

\end{itemize}   

We now come to the particular case when $\beta =0$, that means, the model $\Lambda (H) = 3 H^2 + \delta H^n$ In this case, one can solve the evolution equation (\ref{bernoulli}) leading to

\begin{eqnarray}
H = \Bigl [H_0^{1-n} + \frac{\gamma \delta}{2} (1-n) (t-t_0) \Bigr]^{\frac{1}{1-n}},
\end{eqnarray} 
and consequently, the scale factor is solved as

\begin{align}\label{bernoulli-beta=0}
\frac{a}{a_0} = \exp \left(-\frac{2 H_0^{2-n}}{(2-n)\gamma \delta} \right) \times  \nonumber\\  \times \exp\left[ \frac{2}{(2-n) \gamma \delta} \left( H_0^{1-n} + \frac{\gamma\delta}{2} (1-n)(t-t_0)\right)^{\frac{2-n}{1-n}} \right]
\end{align}
where we must have $n \neq 2$. For $n =2$, the model is, $\Lambda (H) = (3+\delta) H^2$. This model has been excluded for various reasons described in \cite{Basilakos:2009wi} (also see \cite{Gomez-Valent:2014rxa}), hence, we do not focus on it. Now, one can see that the nature of the solution (\ref{bernoulli-beta=0}) is highly dependent on the factor $\left(\frac{2-n}{1-n}\right)$. Thus, we investigate the solution (\ref{bernoulli-beta=0}) for different ranges of $n$ as follows. For $n>2$, one can see that the factor $\left(\frac{2-n}{1-n}\right) >0$, hence, the factor 
$\left( H_0^{1-n} + \frac{\gamma\delta}{2} (1-n)(t-t_0)\right)^{\frac{2-n}{1-n}}$ does not diverge for any values of $n>2$. This means we have a singularity free solution. Now we consider the interval $1< n <2$ for which the factor $\left(\frac{2-n}{1-n}\right)$ could be negative and hence the factor $\left( H_0^{1-n} + \frac{\gamma\delta}{2} (1-n)(t-t_0)\right)^{\frac{2-n}{1-n}}$ could diverge at some finite cosmic time. One can find that under this restriction, the scale factor may diverge at 

\begin{eqnarray}\label{bernoulli-singul-beta=0}
t = t_0 + \frac{2}{\gamma \delta (n-1)} H_0^{1-n}. 
\end{eqnarray}
Now, since $n>1$, then the singularity described in equation (\ref{bernoulli-singul-beta=0}) could be a future singularity if $\delta>0$, while on the other hand, for $\delta <0$, one finds that the scale factor diverges at finite time in the past which is unphysical. Thus, we arrive at the conclusion that the model $\Lambda (H) = 3 H^2 + \delta H^n$ with $\delta <0$ and $1< n< 2$ does not fit into our framework.

Finally, we consider the case when $n <1$. In this case,  $\left(\frac{2-n}{1-n}\right) >0$, and thus, equation (\ref{bernoulli-beta=0}) represents a singularity free universe.

\subsection{Model 2: $\Lambda(H, \dot{H}, \ddot{H})= \alpha + \beta H + \delta H^2 + \mu \dot{H}+ \nu \ddot{H}$}

In the previous section we have discussed the cosmological solutions when $\Lambda (t)$ is only dependent only on the Hubble rate and its powers.
We in this section we allow the derivatives of the Hubble rate considering a very general cosmological model with
\begin{align}\label{model-general}	
\Lambda (t) & \equiv \Lambda (H, \dot{H}, \ddot{H}) = \alpha + \beta H + \delta H^2 + \mu \dot{H}+ \nu \ddot{H},
\end{align} 
in which ($\alpha,\, \beta,\, \delta,\, \mu, \, \nu\, \in \mathbb{R}$).  We mention that the parameters ($\alpha,\, \beta,\, \delta,\, \mu, \, \nu$) should be taken in such a way so that $\Lambda (t) > 0$. Now, the master equation (\ref{main-eqn}) thus takes the form 

\begin{align}
(\gamma\, \nu) \ddot{H} - (2-\gamma\, \mu) \dot{H} - \gamma (3- \delta) H^2+ (\beta\, \gamma) H + (\alpha \gamma)  & = 0,
\end{align}
and this in principle can determine the dynamics of the universe. One can see that the 
above equation is invariant under time translations. For that if we select $H =v$ and $dH/dt=W(v)$,
then the differential equation becomes the first order Abel equation of the second type:
$W W_v+A W+ B v^2+ C v+ D=0$, where $W_v = dW/dv$, 
and its solution in general cannot be written in a closed form expression, for more
discussions see \cite{Paliathanasis:2016bkb}. 
Thus, in the following we  we shall consider several models starting from the simplest to the complex one and their physical consequences as well. We note that the models introduced below are new $\Lambda (t)$ with some interesting results.

\subsubsection{The Model $\Lambda(\dot{H}) = \mu \dot{H}$}
\label{sec-model-simplest}

We consider the first simple model in this series as $\Lambda(t) = \mu \dot{H}$, with the choice of the parameters $\alpha=\beta = \delta= \nu = 0$, $\mu \neq 0$ in the general model (\ref{model-general}). The differential equation (\ref{main-eqn}) for this model takes the form 
\begin{eqnarray}
(2-\gamma \mu) \dot{H}+ 3 \gamma H^2= 0,
\end{eqnarray}
where $(2-\gamma \mu)\neq 0$, otherwise $(2-\gamma \mu) =0 $ leads to $H =0$ which is unphysical. The solution for the Hubble parameter becomes 

\begin{eqnarray}
H = \frac{H_0 (2-\gamma \mu)}{(2-\gamma \mu)+ 3 \gamma H_0 (t-t_0)},\nonumber
\end{eqnarray}	
and consequently, the scale factor is solved into 

\begin{eqnarray}\label{SF-A1}
\frac{a}{a_0} = \Bigg|1+ \frac{3\gamma H_0}{(2-\gamma \mu)} (t-t_0)\Bigg|^{(2-\gamma \mu)/(3\gamma)},
\end{eqnarray}
which is the power law expansion of the universe.
Now, the solution for the scale factor in (\ref{SF-A1}) may lead to different possibilities depending on the sign of $(2-\gamma \mu)$. For instance, if $(2-\gamma \mu) > 0$, then one can readily see that
the scale factor (\ref{SF-A1}) predicts a finite time past singularity at $t_p = t_0 - \frac{(2-\gamma \mu)}{3\gamma H_0} < t_0$. While on the other hand, for $(2-\gamma \mu) < 0$, the scale factor diverges at some finite future time given by $ t_f = t_0 - \frac{(2-\gamma \mu)}{3\gamma H_0} > t_0$.

\subsubsection{The Model $\Lambda(\dot{H}) = \alpha + \mu \dot{H}$}
\label{sec-model-simplest-extn}

We include a constant to the previous model in section \ref{sec-model-simplest} 
as $\Lambda(H) = \alpha + \mu \dot{H}$. For such model one can write down the 
master equation as 

\begin{eqnarray}\label{ssss}
(2-\gamma \mu) \dot{H} = \gamma (\alpha - 3 H^2).
\end{eqnarray} 
Now one may explore that for $(2-\gamma \mu) = 0$, a special case, the Hubble 
rate becomes constant as $H^2 = \frac{\alpha}{3}$ ($\alpha > 0$) leading to
an exponential expansion of the universe. So, we go for a general solution 
with the condition $(2-\gamma \mu) \neq 0$.  Since $\alpha $ is unrestricted, 
thus we consider two separate cases, namely, $\alpha < 0$ and $\alpha >0$. 
Now, solving the equation (\ref{ssss}) for $\alpha < 0$, one finds that 
the Hubble factor evolves as 

\begin{eqnarray}
H = \left(-\frac{\alpha}{3}\right)^{1/2}~ \tan \Bigl(  A (t-t_0) + B  \Bigr),\nonumber
\end{eqnarray}
where $A = -(-\, \alpha/3 )^{1/2} \left(\frac{3\gamma}{2-\gamma \mu}  \right)$ and $B = \arctan \left(H_0 \left(-\frac{\alpha}{3}\right)^{1/2} \right) > 0$ as $H_0 > 0$. The scale factor is can be solved as

\begin{eqnarray}\label{xsxs}
a = a_0 ~\Bigg| \frac{\cos \left(  A (t-t_0) + B   \right)}{\cos B}   \Bigg|^{-\frac{\alpha \gamma}{2-\gamma \mu}},
\end{eqnarray}
where $\cos B \neq 0$ as one can see using the value of $B$ given above. Now, since $\alpha < 0$, then for $(2-\gamma \mu )> 0$, the solution for the scale factor represents an oscillating universe.  This solution has been found in the context of inhomogeneous equation of state with a more generalized model including the square of $\dot{H}$ in addition to the present terms, see \cite{Nojiri:2005sr, Capozziello:2005pa} for a detailed discussions. However, for $(2-\gamma \mu )< 0$, we find that the scale factor diverges at $t= t_0 + \frac{1}{A} [(2n+1)\frac{\pi}{2} -B]$ ($n \in \mathbb{Z}$). This divergence may occur in the past or future depending on the parameters involved. Although the divergence in the scale factor at finite future time could be interesting but its divergence in the past is not sound from the physical context. 

Now we consider that $\alpha >0$. In this case, the Hubble rate can be solved into

\begin{eqnarray}
H = \xi \left [   \frac{\left( \frac{H_0 +\xi}{H_0 - \xi} \right) \exp \left(A (t-t_0)  \right) +1}{\left( \frac{H_0 +\xi}{H_0 - \xi} \right) \exp \left(A (t-t_0)  \right) -1}    \right],\nonumber
\end{eqnarray}
where $\xi = \sqrt{\alpha/3}$~, $A = \frac{6 \xi \gamma}{2-\gamma \mu}$. The scale factor can be calculated as

\begin{align}\label{pan-new1}
a &= a_0  \left(\frac{H_0-\xi}{2\xi} \right)^{\frac{2\xi}{A}}\, e^{-\xi (t-t_0)} ~ \Bigg| \left( \frac{H_0 +\xi}{H_0 - \xi}\right) e^{A (t-t_0)} -1  \Bigg|^{\frac{2\xi}{A}}.
\end{align}
Now one can see that the scale factor (\ref{pan-new1}) could be zero at some finite time or it could diverge depending on the sign of $A$, equivalently on the sign of $(2-\gamma \mu)$. So, if $A > 0$ (or, $(2-\gamma \mu) > 0$) then the scale factor becomes zero at 

\begin{align}
t_{\ast} = t_0 + \frac{1}{A} \ln \left(\frac{H_0 -\xi}{H_0 +\xi}\right) = t_0 + \frac{2-\gamma \mu}{6 \xi \gamma} \ln \left|\frac{H_0 -\xi}{H_0 +\xi}\right| >t_0,\nonumber
\end{align}
which means that in future the scale factor will vanish, 
while for $A < 0$ (i.e. when $(2-\gamma \mu) < 0$), the scale factor diverges at $t_{\ast} < t_0$, which is of course not physical. That means the most physical scenario is the case one which predicts a future singularity.

\subsubsection{The Model $\Lambda(H, \dot{H}) = \delta H^2 + \mu \dot{H}$}
\label{sec-MM}
One can generalize the model in section \ref{sec-model-simplest} with a new term as $\Lambda(H) =  \delta H^2 + \mu \dot{H}$. In this case, the governing differential equation takes the form 

\begin{eqnarray}\label{pan1}
(2-\gamma \mu ) \dot{H} + \gamma (3 -\delta) H^2 = 0.
\end{eqnarray}
Now, one can solve (\ref{pan1}) for $\delta = 3$ and $\delta \neq 3$. 
For the first case, we find $\dot{H} = 0$ for $(2-\gamma \mu) \neq 0$ which implies that 
$H = $ constant, that means the case for $\delta =3$ gives an exponential 
expansion of the universe. 
Now, concerning the case $\delta \neq 3$, we have 

\begin{eqnarray}
H = \frac{H_0 (2-\gamma \mu)}{(2-\gamma \mu)+ (3-\delta) \gamma H_0 (t-t_0)},\, (\delta \neq 3),\nonumber
\end{eqnarray}	
and 
\begin{eqnarray}
\frac{a}{a_0} = \Bigg|1+ \frac{(3-\delta)\gamma H_0}{(2-\gamma \mu)} (t-t_0)\Bigg|^{\frac{(2-\gamma \mu)}{(3-\delta)\gamma}},\, (\delta \neq 3),\nonumber
\end{eqnarray}
which represents the power law expansion of the universe and alreday obtained for the model $\Lambda (t) = \delta \dot{H}$. Consequently, one can easily verify that the model may  predict cosmic singularities at finite time depending on the signs of both $(3-\delta)$ and $(2-\gamma \mu)$. \textit{Let us consider Case I:} Here, we first assume that $\delta < 3$. In this case we could have two different possibilities depending on the sign of $(2-\gamma \nu)$: For $(2-\gamma \mu) > 0$, a finite time singularity at past is detected at $t_p= t_0 -\frac{(2-\gamma \mu)}{\gamma(3-\delta)H_0} <t_0$, while for $(2-\gamma \mu) < 0$, the scale factor diverges at some finite time in future given by $t_f = t_0 - \frac{(2-\gamma \mu)}{\gamma (3-\delta)H_0} > t_0$. \textit{Now we consider Case II:} In this case we assume  that $\delta > 3$. One can find that depending on the sign of $(2-\gamma \mu)$ two different phases may occur. For $(2-\gamma \mu) > 0$, the scale factor diverges at some finite time in future given by $t_f = t_0 -\frac{(2-\gamma \mu)}{\gamma (3-\delta) H_0} > t_0$. Now, on the other hand, for $(2-\gamma \mu) < 0$, finite time past singularity is observed at  
$t_p = t_0 -\frac{(2-\gamma \mu)}{\gamma(3-\delta)H_0} < t_0$. 
Thus, from the solutions one can clearly see that the future singularity can be avoided under certain restrictions on the model parameters where the conditions are either (a) $\delta <3$ and $(2-\gamma \mu) >0$ or (b) $\delta >3$ and $(2-\gamma \mu) < 0$. The equivalent cosmological scenrio in terms of the inhomogeneous equation of state has been studied in \cite{Capozziello:2005pa, Bamba:2012cp} where the restrictions on the models to avoid any future singularity have been discussed.

\subsubsection{The model $\Lambda (H, \dot{H}) = \delta H^2 + \mu \dot{H} +\alpha$}
\label{sec-MMM}
We consider another model where $\Lambda (t) = \delta H^2 + \mu \dot{H} +\alpha$ ($\alpha \neq 0$). This is an extension of the model in section \ref{sec-MM}. We aim to see the changes in the cosmic evolution when a nonzero constant is added. Now, 
for this model the master differential equation is 

\begin{eqnarray}\label{mmm}
(2-\gamma \mu) \dot{H} = \gamma \left[ \alpha - (3-\delta) H^2 \right].
\end{eqnarray}
Now, from (\ref{mmm}) one can derive several conclusions as follows. Let us first consider 
the case when $(2-\gamma \mu) = 0$, this choice leads to the solution $H^2 = \frac{\alpha}{3-\delta}$, which is defined for either ($\delta < 3$ and $\alpha > 0$) or ($\delta > 3$ and $\alpha <0$). Thus, it is strainghtforward that the scale factor for this choice leads to an exponential expansion of the universe. Now we move on to the next possibility when  $(2-\gamma \mu) \neq  0$ and $\delta = 3$. This choice leads to the following solutions. The Hubble parameter evolves as $H = H_0 + \frac{\alpha}{2-\gamma \mu} (t-t_0)$, and the scale factor evolves as $a = a_0 \exp \left[ H_0 (t-t_0) + \frac{\alpha}{2 (2-\gamma \mu)} (t-t_0)^2\right]$, which again shows an exponential expansion, but we note that the evolution of the Hubble factor is different from the first case. Now we move on to the more general case when  $(2-\gamma \mu) \neq  0$ and $\delta \neq 3$.  Under these conditions, we arrive at the following integral

\begin{eqnarray}
\int_{H_0}^ {H} \frac{dH}{\left( \frac{\alpha}{3-\delta} \right)-H^2} = \frac{\gamma (3-\delta)}{2-\gamma \mu} (t-t_0),
\end{eqnarray}
which depending on the sign of $\left( \frac{\alpha}{3-\delta} \right)$ may lead to different solutions for the Hubble rate. We consider the first case when this quantity assumes negative values, i.e. $\left( \frac{\alpha}{3-\delta} \right) = - v^2 <0$, where $v>0$. The Hubble rate can be solved into 

\begin{eqnarray}
H = v \tan \Bigl[ A + B (t-t_0) \Bigr],
\end{eqnarray} 
where $A = \tan^{-1} (H_0/v)$ and $B = - \frac{v\gamma (3-\delta)}{(2-\gamma \mu)}$. Consequently, the evolution of the scale factor becomes,

\begin{eqnarray}
\frac{a}{a_0} = \Bigg| \frac{\cos \left( A + B (t-t_0)\right)}{\cos A}\Bigg|^{\frac{2-\gamma \mu}{\gamma (3-\delta)}},
\end{eqnarray}
where $\cos A \neq 0$, as one can easily find using the expression for $A$ given above. One may observe that two different conditions may arise out of this solution. For $\frac{2-\gamma \mu}{\gamma (3-\delta)} >0$, we have an oscillating universe. We note that an equivalent model of $\Lambda (t)$ in terms of an inhomogeneous equation of state described in \cite{Nojiri:2005sr, Capozziello:2005pa} exhibits the oscillating universe scenario. While on the other hand, for $\frac{2-\gamma \mu}{\gamma (3-\delta)}<0$, the scale factor can diverge at finite time given by

\begin{eqnarray}
t =t_0 - \frac{2-\mu \gamma}{v(3-\delta)\gamma} \Bigl[ (2n+1) \frac{\pi}{2} - \tan^{-1}(H_0/v)\Bigr]
\end{eqnarray}
where $n \in \mathbb{Z}$. Since $\frac{2-\gamma \mu}{\gamma (3-\delta)}<0$, and $v$ is a positive quantity, thus, the scale factor can diverge at finite time in future if $(2n+1) \frac{\pi}{2} - \tan^{-1}(H_0/v) >0$ while from the mathematical interest, one can see the scale factor will diverge at some finite comsic time in past provided, $(2n+1) \frac{\pi}{2} - \tan^{-1}(H_0/v) <0$. 

We now focus on the second possibility when the quantity $\left( \frac{\alpha}{3-\delta} \right)$ is positive. Let's say $\left( \frac{\alpha}{3-\delta} \right) = p^2$ ($p> 0$). 
Henceforth, the Hubble rate can be solved as 

\begin{eqnarray}\label{new-added1-sp}
H = p \tanh \left[ \tanh^{-1} \left(\frac{H_0}{p} \right) + \frac{p\gamma (3-\delta)}{2-\mu \gamma} (t-t_0) \right],
\end{eqnarray}
for $H < p$ while for $H >p$, this becomes

\begin{eqnarray}\label{new-added2-sp}
H = p \coth \left[ \coth^{-1} \left(\frac{H_0}{p} \right) + \frac{p\gamma (3-\delta)}{2-\mu \gamma} (t-t_0) \right],
\end{eqnarray}

Now one can find the exact evolution equations for the scale factors using the above equations for the Hubble factors and they are

\begin{equation}
\frac{a}{a_0} = \Bigg[\frac{\cosh \left( \tanh^{-1} \left(\frac{H_0}{p} \right) + \frac{p\gamma (3-\delta)}{2-\mu \gamma} (t-t_0) \right)}{\cosh A}\Bigg]^{\frac{2-\mu \gamma}{\gamma (3-\delta)}},
\end{equation} 
for (\ref{new-added1-sp}) which clearly represents a singularity free solution and for (\ref{new-added2-sp}), one has

\begin{eqnarray}\label{new-added3-sp}
\frac{a}{a_0} =  \Bigg| \frac{\sinh \left[ \coth^{-1} \left(\frac{H_0}{p} \right) + \frac{p\gamma (3-\delta)}{2-\mu \gamma} (t-t_0)\right]}{\sinh \left( \coth^{-1} \left(\frac{H_0}{p} \right) \right)} \Bigg|^{\frac{2-\mu \gamma}{\gamma (3-\delta)}},
\end{eqnarray}
which may lead to finite time singularities in the scale factor. We note that $\sinh \left( \coth^{-1} \left(\frac{H_0}{p} \right) \right) \neq 0$. Now let us investigate this solution closely. The index $\frac{2-\mu \gamma}{\gamma (3-\delta)}$ could be both positive and negative. We consider the positive case firstly. In that case, the scale factor could vanish at some finite cosmic time given by 

\begin{eqnarray}
t = \left[ t_0 - \frac{1}{p \gamma} \left(\frac{2-\mu \gamma}{3-\delta}\right) \coth^{-1} (H_0/p)\right] < t_0
\end{eqnarray} 
since the quantity $\frac{2-\mu \gamma}{\gamma (3-\delta)}>0$ and hence the finite time singularity at past is realized. Now on the other hand, for $\frac{2-\mu \gamma}{\gamma (3-\delta)} <0$, one can realize the future  singularity at the finite time given by 

\begin{eqnarray}
t = \left[ t_0 - \frac{1}{p \gamma} \left(\frac{2-\mu \gamma}{3-\delta}\right) \coth^{-1} (H_0/p)\right] > t_0
\end{eqnarray} 
since $\frac{2-\mu \gamma}{\gamma (3-\delta)} <0$. We note that with suitable choices for the free parameters the big-rip singularity can be avoided, that means with the choice of $\frac{2-\mu \gamma}{\gamma (3-\delta)}>0$, there is no finite time future singularity.

\subsubsection{Model: $\Lambda (H, \dot{H}, \ddot{H})=  3H^2 + \mu \dot{H}+ \nu \ddot{H}$}
\label{sec-xsxsxsxs}

We consider the model $\Lambda (H, \dot{H})= 3 H^2 + \mu \dot{H}+ \nu \ddot{H}$, with the choices of the free parameters $\alpha = \beta = 0$, $\delta = 3$. For this model, the master differential equation (\ref{main-eqn}) becomes

\begin{align}
\gamma\, \nu \, \ddot{H} - (2-\gamma\, \mu)\, \dot{H} & = 0.
\end{align}
We can solve this differential equation leading to the Hubble factor as well as the scale factor, respectively as,

\begin{align}
H  = H_0 \exp \left[ \left(\frac{2-\gamma \mu}{\gamma \nu}\right) \left(t-t_0  \right)\right],\nonumber
\end{align}
and 

\begin{eqnarray}
\frac{a}{a_0}  = \exp \left[ H_0 \left( \frac{\gamma \nu}{2-\gamma \mu}\right) \left(e^{\left(\frac{2-\gamma \mu}{\gamma \nu}\right)(t-t_0)}-1\right) \right].\nonumber
\end{eqnarray}
This solution for the scale factor is of double exponential type which cannot be
zero at any finite time, hence no singularity during the  past or future evolution 
of the universe appears.

\subsection{Model: $\Lambda (H, \dot{H}, \ddot{H})=  \alpha + 3H^2 + \mu \dot{H}+ \nu \ddot{H}$}

This model is the generalization of the previous model discussed in section \ref{sec-xsxsxsxs}. The master equation (\ref{main-eqn}) takes the form

\begin{eqnarray}
\gamma \nu \ddot{H} - (2-\gamma \mu)\dot{H} + \alpha \gamma = 0~,
\end{eqnarray}
which describes the evolution of the universe. Let us consider the case when $(2-\gamma \mu) = 0$, which is nothing but the model $\Lambda (H, \dot{H}, \ddot{H})=  \alpha + 3H^2 + \left(\frac{2}{\gamma} \right) \dot{H}+ \nu \ddot{H}$. One can see that for such particular model, the Hubble rate can be solved as 
\begin{eqnarray}
H = H_0 -  \left(\frac{\alpha}{2\,\nu}\right)\, \left(t-t_0 \right)^2 - \left(1+q_0 \right)\, H_0^2\, (t-t_0),\nonumber
\end{eqnarray}
where $q_0$ is the present value of the deceleration parameter of the FLRW universe defined by, $q = -1-\frac{\dot{H}}{H^2}$. The solution for the scale factor can be immediately solved from the above equation that gives 

\begin{eqnarray}
\ln \left(\frac{a}{a_0}\right) = H_0 (t-t_0)- \frac{\alpha}{6 \nu} (t-t_0)^3  - \frac{H_0^2}{2}(1+q_0) (t-t_0)^2,\nonumber
\end{eqnarray}
which is a singualrity free solution and describes an exponential expansion of the universe. \newline

Now let us consider the case when $(2-\gamma \mu )\neq 0$. In this case, the Hubble parameter  assumes the closed form solution  as

\begin{eqnarray}
H = A_c + B_c \exp \left[ \left( \frac{2-\gamma \mu}{\gamma \nu} \right)t \right]  + \frac{\alpha \gamma}{2-\gamma \mu} \left( t+ \frac{\gamma \nu}{2-\gamma \mu} \right),\nonumber
\end{eqnarray}
where $A_c$, $B_c$ are the integration constants. Certainly, from the solution of the Hubble rate, it is clearly seen that this solution also represents a singualrity free universe and the expansion of the universe is exponential.

\subsubsection{The model $\Lambda (H, \dot{H}, \ddot{H})= \beta H + 3 H^2 + \mu \dot{H}+ \nu \ddot{H}$}

We consider now a model with with $\delta = 3$, $\alpha =0$. That means, 
$\Lambda (H)= \beta H + 3 H^2 + \mu \dot{H}+ \nu \ddot{H}$. 
This leads to the following differential equation

\begin{align}\label{eqn1}
(\gamma\, \nu)\, \ddot{H} - (2-\gamma\, \mu)\, \dot{H} + (\beta\, \gamma)\, H   & = 0
\end{align}
which is a second order linear differential equation with constant coefficients. The roots of the auxiliary equation of the differential equation (\ref{eqn1}) are

\begin{align}
m_1 & = \frac{(2-\gamma\, \mu)+ \sqrt{(2-\gamma \mu)^2 -4 \gamma^2\,\beta\,\nu}}{2\gamma \, \nu}\\
m_2 & = \frac{(2-\gamma\, \mu)- \sqrt{(2-\gamma \mu)^2 -4 \gamma^2\,\beta\,\nu}}{2\gamma \, \nu}
\end{align}
Now, depending on $\Delta= (2-\gamma \mu)^2 -4 \gamma^2\,\beta\,\nu$, we could have three different
solutions. \newline

$\bullet$ When $\Delta> 0$, that means, both $m_1$ and $m_2$ are real and unequal, we have the following solution of the Hubble
parameter

\begin{align}\label{hubble1}
H & = C_1\, e^{m_1\, t} + C_2\, e^{m_2\, t},
\end{align}
where $C_1$, $C_2$ are the arbitrary constants. Solving the equation (\ref{hubble1}), one can get the scale factor as

\begin{align}\label{SF-A2}
\frac{a}{a_0} = \exp \left[ \frac{C_1}{m_1} \left( e^{m_1 t} - e^{m_1 t_0}\right)+  \frac{C_2}{m_2} \left( e^{m_2 t} - e^{m_2 t_0}\right)\right],
\end{align}
which clearly presents a universe with no singularity and the evolution law is double exponential type. \newline

$\bullet$ Now when $\Delta = (2-\gamma \mu)^2 -4 \gamma^2\,\beta\,\nu = 0$, that means both the roots
are equal, i.e. $m_1=m_2= m = (2-\gamma \mu)/2\gamma \nu$, then the solution of the Hubble parameter becomes

\begin{align}
H & = (D_1 + D_2\, t)\, e^{m\,t},\nonumber
\end{align}
where $D_1$, $D_2$, are all integration constants. Now, one can solve the expansion scale factor as

\begin{align}
\frac{a}{a_0} = \exp \Bigg[\frac{e^{mt_0}}{m} \left(D_1 - \frac{D_2}{m} \right) \left(e^{m(t-t_0)}-1\right) \nonumber\\+ \frac{D_2 e^{mt_0}}{m} \left( t e^{m(t-t_0)} -t_0\right)\Bigg],\nonumber
\end{align} 
which describes a nonsingular universe and in addition the scale factor is double exponential. \newline

$\bullet$ Finally, we consider the case when $\Delta = (2-\gamma \mu)^2 -4 \gamma^2\,\beta\,\nu < 0$, that means this case
will lead to two complex roots $r = r_1 \, \pm \, i r_2$ ($i= \sqrt{-1}$), where
$r_1= \frac{(2-\gamma\, \mu)}{2\gamma \, \nu}$, and $r_2 = \frac{\sqrt{4 \gamma^2\,\beta\,\nu- (2-\gamma \mu)^2}}{2\gamma \, \nu}$.
Thus, in this case, the solution for the Hubble parameter becomes

\begin{align}
H & = \exp\left(r_1\, t \right)\, \Bigl[ E_1\, \cos(r_2\, t)+ E_2\, \sin(r_2\, t) \Bigr],\nonumber
\end{align}
where $E_1$, $E_2$ are arbitrary constants. This solution also leads to an oscillating scale factor. \newline 

We consider a special case of the model $\Lambda(H, \dot{H}, \ddot{H})= \beta H + 3 H^2 + \mu \dot{H}+ \nu \ddot{H}$, where $\mu= 2/\gamma$. Although one can treat this to be a purely academic choice, 
but however, here we are interested in the simplicity of the cosmological evolutions.  
The Raychaudhuri equation for this model turns out to be
\begin{eqnarray}
\nu \ddot{H}+ \beta\, H = 0, 
\end{eqnarray}
which is a second order
differential equation with constant coefficients. The auxiliary equation becomes
$m^2+ \beta/\nu = 0$. Now, depending on the signs of $\beta/\nu$, we will have three
different possibilities as follows:\newline

$\bullet$ When $\beta/\nu < 0$, that means either ($\beta> 0$, $\nu < 0$) or ($\beta< 0$, $\nu > 0$), the roots
of the auxiliary equation becomes real and unequal as $r_1 = \sqrt{-\frac{\beta}{\nu}}$, $r_2= - \sqrt{-\frac{\beta}{\nu}}$,
and thus the solution for the Hubble parameter takes the form

\begin{align}
H & = f_1\, \exp\left(r_1 \,\,t\right) + f_2 \exp\left(r_2 \,\,t\right),\nonumber
\end{align}
where $f_1$, $f_2$ are any two arbitrary integration constants. Now, the scale
factor in this case becomes

\begin{align}
\ln \left(\frac{a}{a_0}  \right)= \frac{f_1}{r_1} e^{r_1\,t_0} \left ( \exp(r_1 (t-t_0)) -1 \right) \nonumber\\+ \frac{f_2}{r_2} e^{r_2\,t_0} \left ( \exp(r_2 (t-t_0)) -1 \right),\nonumber
\end{align}
which shows that the scale factor does not have any finite time singularity during the evolution of the universe, and moreover, the expansion scale factor is of double exponential type. \newline

$\bullet$ The roots of the auxiliary equation will be equal when $\beta = 0$, so in this case the solution for the Hubble parameter becomes

\begin{align}
H & = g_1 + g_2\, t,\nonumber
\end{align}
where $g_1$, $g_2$ are arbitrary constants. Using the condition  at $t= t_0$, $H = H_0$, the above equation for the Hubble factor can be written as $H = H_0 + g_2 (t-t_0)$, which on integration solves the  scale factor as

\begin{align}
a & = a_0 \, \exp\left[ H_0 \, ( t-t_0 ) + \frac{g_2}{2}\left(t-t_0\right)^2 \right],\nonumber
\end{align}
which again shows a singularity free universe with an exponential expansion. \newline

$\bullet$ Finally, if we have either ($\beta> 0$, $\nu > 0$) or ($\beta< 0$, $\nu < 0$), then the roots of $m^2+ \beta/\nu = 0$, are purely imaginary. We denote the roots by $m = \pm~ i ~ \theta$, where $\theta = \sqrt{\frac{\beta}{\nu}}$. The explicit solution for the Hubble parameter becomes

\begin{align}
H & = h_1\, \cos\left(\theta t\right) + h_2\, \sin \left(\theta t \right),\nonumber
\end{align}
where $h_1$, $h_2$ are the constants of integrations. The scale factor this time takes the following
form

\begin{align}
\ln \left(  \frac{a}{a_0}  \right) = \frac{h_1}{\theta} \Bigl[ \sin (\theta t) - \sin (\theta t_0)  \Bigr]- \frac{h_2}{\theta} \Bigl[ \cos (\theta t) - \cos (\theta t_0)  \Bigr]\nonumber.
\end{align}
This represents an oscillating universe model.
Finally, we note that one can generalize this scenario with the model 
$\Lambda (H)= \alpha + \beta H + 3 H^2 + \mu \dot{H}+ \nu \ddot{H}$
for which the master equation turns out to be 
\begin{align}\label{nice-equation}
(\gamma\, \nu)\, \ddot{H} - (2-\gamma\, \mu)\, \dot{H} + (\beta\, \gamma)\, H + (\alpha\, \gamma)  & = 0,
\end{align}
and  following the same procedure described above one can explore the physics behind the solutions.

\section{Summary and Conclusions}
\label{discu}

The cosmology with time dependent $\Lambda$ is not a new 
theory \cite{Ozer:1985wr,Peebles:1987ek,Wetterich:1994bg}. 
This theory became more popular at recent time when the rigid
cosmological constant was questioned by recent observations. 
Thus so far, along with other alternative cosmological models, 
the theory of $\Lambda (t)$ gained much attention in the 
scientific community \cite{BC, CPBA, CDPA, PDCA, Basilakos:2009wi, Bauer:2009ea, Lima:2012mu, Perico:2013mna, Basilakos:2013xpa, Gomez-Valent:2014rxa, Gomez-Valent:2014fda, Gomez-Valent:2015pia,Oikonomou:2016pnq} (also see \cite{Sola:2013gha}). A quantum 
field theoretic development in this direction 
\cite{Sola:2007sv,Shapiro:2009dh,Sola:2013gha,Sola:2015csa}
fueled such investigations and inspired many cosmologists
to extract information out of the phenomenological
$\Lambda (t)$ models. Although such models do not
have any covariant formulation but an equivalent foundation
in terms of modified gravity (also applicable for the Einstein gravity 
as a special case) generating an 
inhomogeneous equation 
of state can be made \cite{Nojiri:2005sr, Nojiri:2006ri, Capozziello:2005pa,  Nojiri:2010wj}, and thus, for any specific $\Lambda (t)$ model, an equivalent description in terms of some inhomogeneous equations of state can be associated. For the cosmological solutions with such inhomogeneous equations of state, we refer to \cite{Nojiri:2005sr, Nojiri:2006ri, Capozziello:2005pa,  Nojiri:2010wj}.

Thus, being motivated, in this work we consider a class of  $\Lambda (t)$ cosmologies 
in a spatially flat FLRW universe where $\Lambda (t)$ 
has been taken to be the function of the FLRW Hubble 
rate $H$ and its cosmic time derivatives. 
We cover a series of $\Lambda (t)$ (equivalently, $\Lambda (H,\dot{H}, \ddot{H})$) 
models in order to see their dynamical evolutions and viabilities
and specifically we focus on the finite time cosmological singularities
allowed by the models. 
The models in this series have been found to solve the 
evolution equations analytically that lead
to the closed form solutions for the Hubble 
rate and the expansion scale factor $a (t)$.
The solutions for the scale factor from different models include power law, 
exponential, double exponential and oscillating universe models.
Similar type solutions have been found in the context 
of  a cosmological scenario with an inhomogeneous equation of state 
\cite{Nojiri:2005sr, Nojiri:2006ri, Capozziello:2005pa, Nojiri:2010wj, Bamba:2012cp}.
Further, we found that some of the models may allow finite time 
singularities in past or future depending on the free 
parameters of the models. Although under certain conditions, 
the future singularities can be 
avoided in some models. Such observations,
specifically, the future singularity is important 
for any cosmological model.   
Additionally, and interestingly, some of the models allow a singularity free scale factor. The realization of cyclic universe in this context \cite{Pavlovic:2017umo} is a new addition in the literature.

Thus, the time varying Lambda cosmologies offering some interesting and well known expansion laws for the universe together with the finite time singularities in past or future actually impose the theoretical limitations on the models. Finally, we close the work with a comment that the observational estimations of the model parameters will be more realistic, mainly to arrive at a decisive conclusion on the viabilities of the models and toward a more sound explanations.\\\\

\section*{Acknowledgments}
I thank the referee for raising some important issues and helpful comments to improve the work.
This article is financially  supported by the SERB$-$NPDF grant (File No. PDF/2015/000640), Govt. of India. Also, I thank Dr Andronikos Paliathanasis and Prof. J. D. Barrow for some important comments and suggestions for this article.

\end{document}